\documentclass{kluwer}     
\usepackage{graphics}
\def\cc{\,{\rm cm^{-3}}}
\def\cm2{\,{\rm cm^{-2}}}
\def\kms{\,{\rm {km\,s^{-1}}}}
\def\kkms{\,{\rm {K\,km\,s^{-1}}}}
\def\h2{\,{\rm H_{2}}}
\def\13co{\,{\rm ^{13}CO}}
\def\co{\,{\rm ^{12}CO}}
\def\pci{\,{\rm ^{3}P_{1}-^{3}P_{0}\,[CI]}}

\def\mu{\,\mu m}
\def\aua{A\&A, }

\def\apj{ApJ, }

\begin{document}

\begin{article}
\begin{opening}         

\title{CO, $\13co$ and [CI] in galaxy centers}
\author{F.\,P. \surname{Israel}}
\runningauthor{F.P. Israel}
\institute{Sterrewacht Leiden, P.O. Box 9513, NL 2300 RA Leiden,
             The Netherlands}
\date{}
 
\begin{abstract}
Measurements of [CI], $J$=2--1 $\13co$ and $J$=4--3 $\co$ emission
from quiescent, starburst and active galaxy centers reveal a distinct
pattern characterized by relatively strong [CI] emission. The [CI] to
$\13co$ emission ratio increases with central [CI] luminosity. It is
lowest in quiescent and mild starburst centers and highest for strong
starburst centers and active nuclei. C$^{\circ}$ abundances are close
to, or even exceed, CO abundances.  The emission is characteristic of
warm and dense gas rather than either hot tenuous or cold very dense
gas.  The relative intensities of CO, [CI], [CII] and far-infrared
emission suggest that the dominant excitation mechanism in galaxy
centers may be different from that in Photon-Dominated Regions (PDRs).
\end{abstract}

\end{opening}
 
\section{Introduction}

%
\begin{figure*}[t]
\unitlength1cm
\begin{minipage}[]{12cm}
\resizebox{11.5cm}{!}{\rotatebox{270}{\includegraphics*{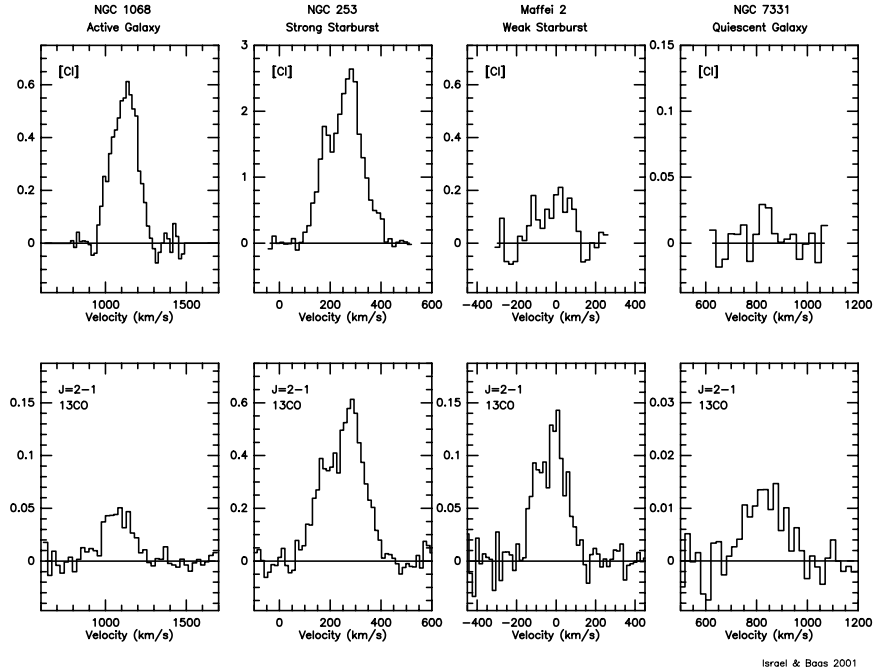}}}
\end{minipage}
\caption[]{[CI] (top) and $J$=2--1 $\13co$ (bottom) spectra observed
towards galaxy centers. The vertical scale is $T_{\rm mb}$ in K; the
horizontal scale is velocity $V_{\rm LSR}$ in $\kms$. Note decreasing
ratio [CI]/$\13co$ going from left to right. }
\label{profiles}
\end{figure*}

Carbon monoxide (CO), the most common molecule after $\h2$, is
routinely detected in external galaxies. When exposed to photons of
energy greater than 11.1 eV, CO is readily photodissociated into
atomic carbon and oxygen. As the C$^{\circ}$ ionization potential is
only 11.3 eV, i.e. quite close to the CO dissociation energy, neutral
carbon may be quickly ionized. Because carbon monoxide, its isotopes,
as well as neutral and ionized carbon respond differently to ambient
conditions, observations of the relative emission strengths of $\co$,
$\13co$, C$^{\circ}$ and C$^{+}$ provide significant information on
the physical condition of cloud complexes from which the emission
arises. Even though far-infrared continuum and [CII] lines are much
more efficient coolants overall, CO and [CI] lines are especially
important for the temperature balance of cool and dense molecular
gas. 

\section{Relative intensities of CO $^{13}$CO and [CI]}

CO may be observed from the ground in many transitions. Emission
from C$^{+}$ has been observed towards numerous galaxies from the KAO
and ISO platforms. Emission from C$^{\circ}$ can also be measured from
the ground but only under excellent atmospheric conditions.  To date,
the total number of galaxies detected in [CI] is about 30.  Most of
the extragalactic 492 GHz [CI] measurements are summarized in just two
papers, by Gerin $\&$ Phillips (2000) using the CSO and by Israel $\&$
Baas (2002) using the JCMT, both on Mauna Kea, Hawaii. These papers
include references to more detailed discussions of individual galaxy
results.

{\em Almost all galaxies mapped thus far show strong concentrations of
both atomic carbon and molecular gas well contained within radii $R
\leq 0.5 $ kpc.} The area-integrated CO and [CI] luminosities of the
observed galaxy centers cover a large range. Quiescent centers
(NGC~7331, IC~342, Maffei~2, NGC~278, NGC~5713) have modest [CI]
luminosities $\approx 1 \leq L_{\rm [CI]} \leq 5 \kkms$ kpc$^{2}$
(with $1 \kkms$kpc$^{2}$ corresponding to $2.2\times10^{20}$ W).
Starburst nuclei (NGC~253, NGC~660, M~82, NGC~3628, NGC~6946) have
luminosities $ 10 \leq L_{\rm [CI]} \leq 40 \kkms$ kpc$^{2}$, except
M~83 which has only $L_{\rm [CI]} = 3.6 \kkms$ kpc$^{2}$. The highest
luminosities, $L_{\rm [CI]} \geq 50 \kkms$ kpc$^{2}$, are found around
the active nuclei of NGC 1068 and NGC 3079.

In previously observed Galactic photon-dominated regions (PDRs), the
intensities of $\pci$ and $J$=2--1 $\13co$ line emission were
generally found to be very similar (cf. Keene et al. 1996; Kaufman et
al. 1999). Such ratios of $\pci$ and $J$=2--1 $\13co$ close to or less
than unity are considered to be characteristic of the effects of
enhanced UV radiation on molecular gas in PDRs.  However,
Fig.\,\ref{profiles} shows that galaxy centers may have much stronger
[CI] emission.

%
\begin{figure*}[t]
\unitlength1cm
\begin{minipage}[b]{6.4cm}
\resizebox{5.5cm}{!}{\rotatebox{270}{\includegraphics*{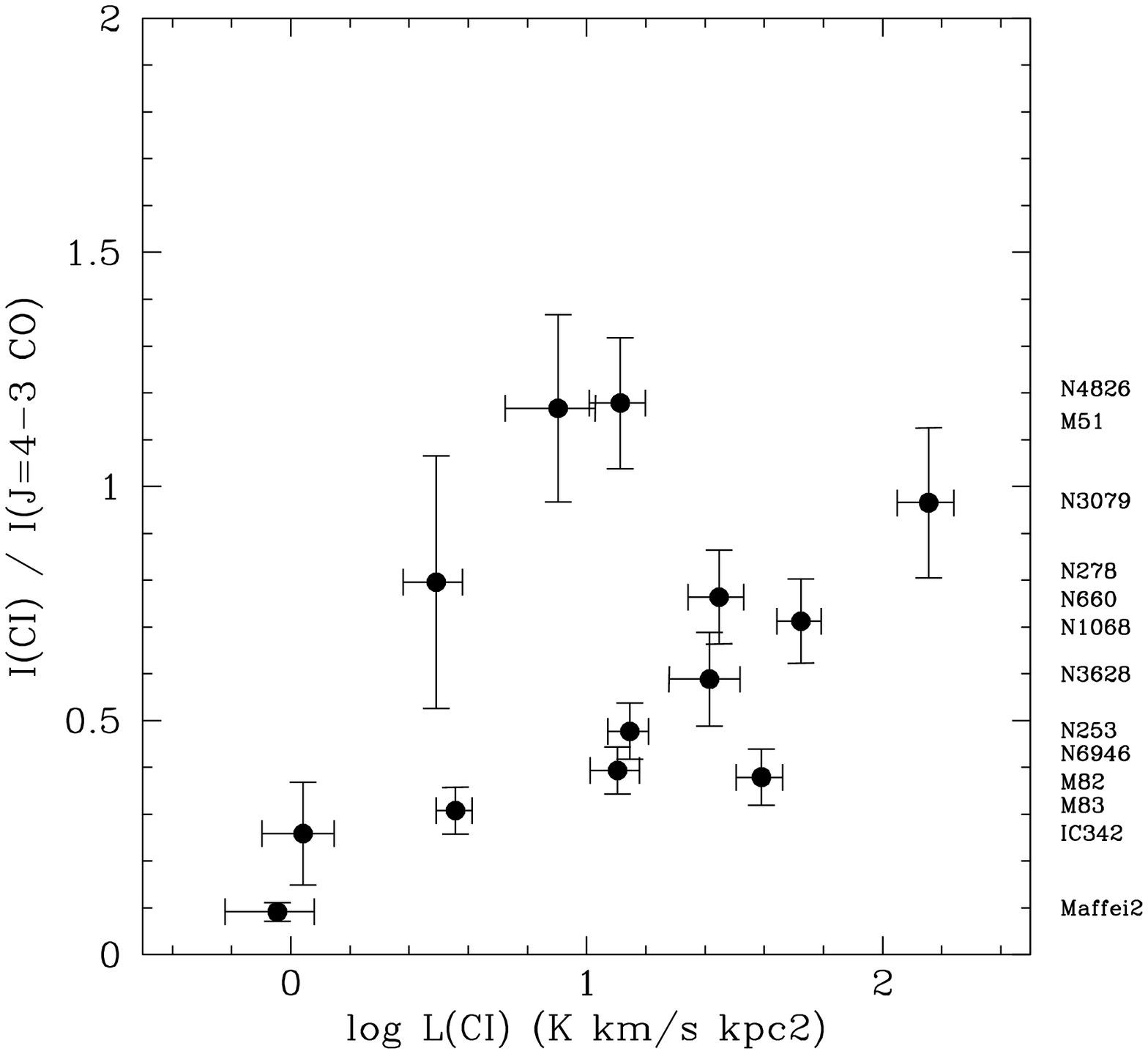}}}
\end{minipage}
\hfill
\begin{minipage}[b]{6.4cm}
\resizebox{5.5cm}{!}{\rotatebox{270}{\includegraphics*{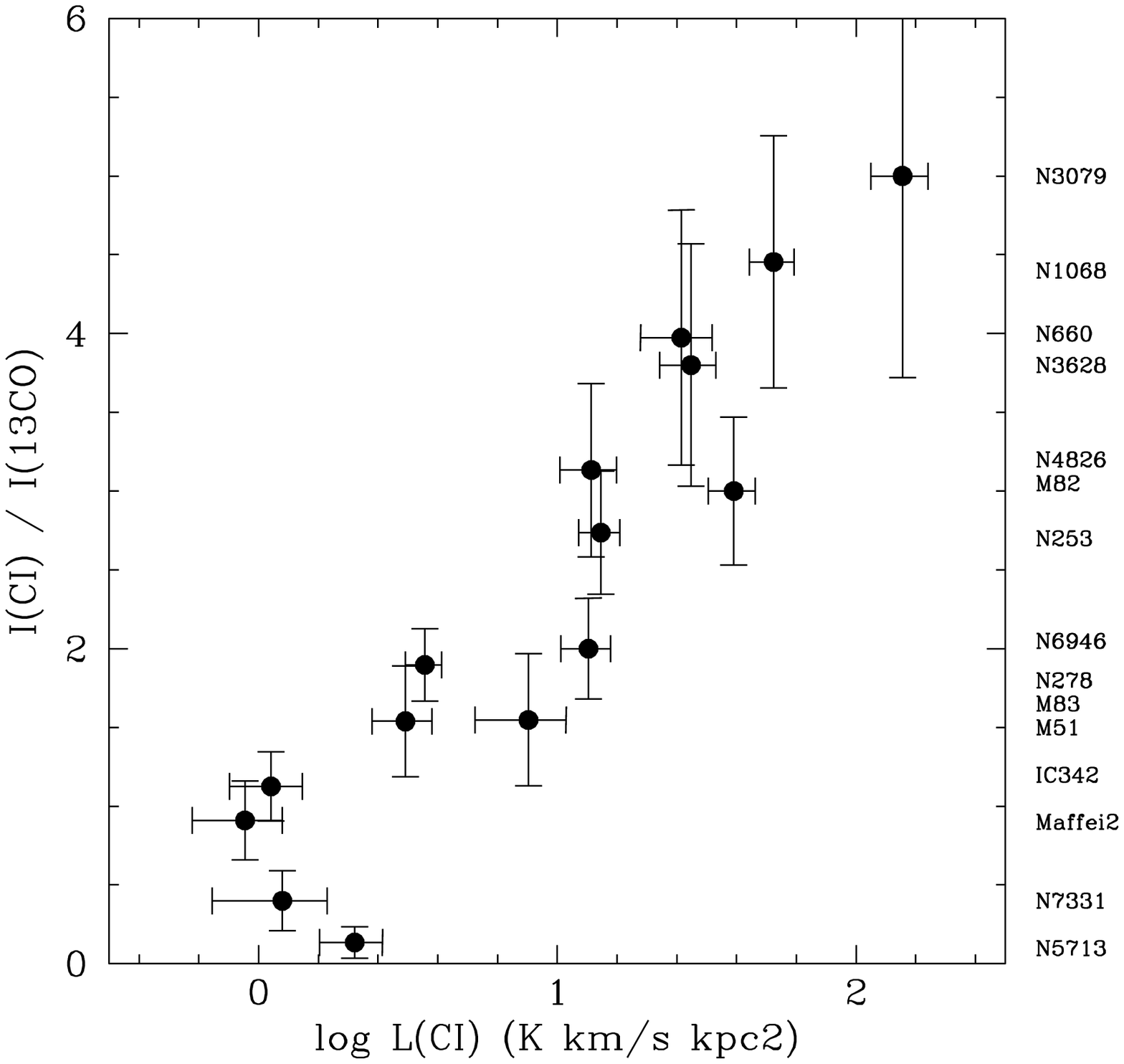}}}
\end{minipage}
\caption[]{Left: [CI]/($J=4-3 \co$) ratios versus area-integrated
luminosity $L[CI]$. Right: [CI]/($J=2-1 \13co$) ratios versus $L[CI]$.
The $I$([CI])/$I(\13co)$ ratio appears to be a well-defined function
of log $L$([CI]); the $I$([CI])/$I$(4--3$\co$) ratio is not. Galactic
sources (not shown) would all be crowded together in the lower left
corner.  }
\label{cilum}
\end{figure*}

In Fig.\,\ref{cilum} we present a full comparison of the intensities of
the 492~GHz [CI] line, the 461~GHz $J=4-3 \co$ line and the 220~GHz
$J$=2--1 $\13co$ line. The $\pci$ line is stronger than $J$=2--1 $\13co$
in all except three galaxies.  The highest [CI]/$\13co$ ratios of
about five belong to the active galaxies NGC~1068 and NGC~3079.
Generally, the $\pci$ line is weaker than the $J$=4--3 $\co$ line, but
not by much. {\em Thus, galaxy centers have much stronger [CI]
emission than the Galactic PDR results would lead us to expect.}

In the galaxy sample independently observed by Gerin $\&$
Phillips (2000) more than two thirds also have a ratio [CI]/$\13co$
$\geq 2$. Low ratios are expected in high-UV environments, and in
environments with high gas (column) densities where virtually all
carbon will be locked up in CO. In the Galaxy, high ratios are found
in environments with low gas column densities and mild UV radiation
fields, such as exemplified by translucent clouds and at cloud
edges. There, $\co$ and especially $\13co$ will be mostly dissociated,
and gas-phase atomic carbon may remain neutral. In dense clouds, [CI]
may be relatively strong if the dominant heating mechanism is some
other than UV photons. The data presented here and in Gerin $\&$
Phillips (2000) suggest that most of {\em the emission from galaxy
centers does not come from very dense, starforming molecular cloud
cores or PDR zones}.

\section{Modelling of [CI], CO and $^{13}$CO}

%
\begin{figure*}[t]
\unitlength1cm
\resizebox{11.8cm}{!}{\rotatebox{270}{\includegraphics*{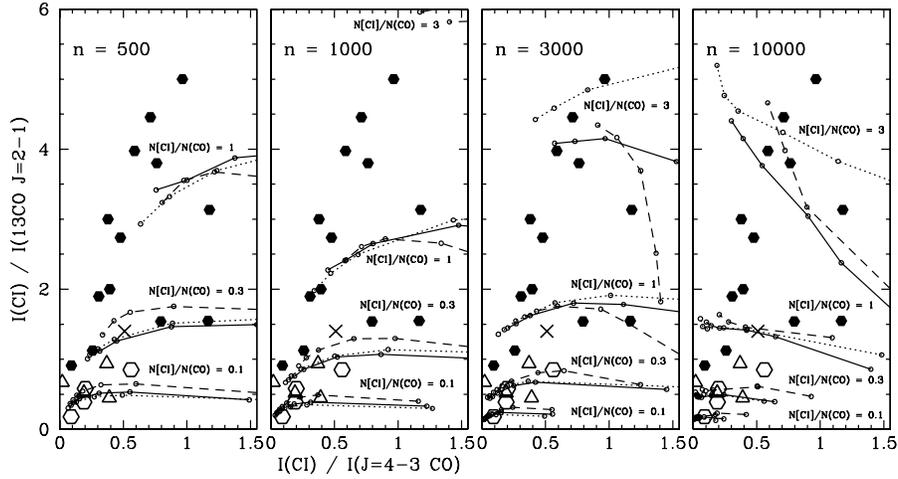}}}
\caption[]{Observed line intensity ratios [CI]/$\13co$ versus [CI]/CO
4--3 compared to LVG radiative-transfer model ratios at selected gas
densities, temperatures and velocity gradients assuming an isotopic
ratio $[\co]/[\13co]$ = 40. Galaxy centers are marked by filled
hexagons, PDRs by open hexagons (LMC) or open triangles (Milky Way),
and the Milky Way Center by a cross. Line families indicate abundances
N[CI]/N(CO) = 0.1, 0.3, 1 and 3. Within each family, a dotted line
corresponds to a gradient $N$(CO)/d$V=3\times10^{16} \cm2/\kms$, solid
line to $N$(CO)/d$V=1\times10^{17} \cm2/\kms$ and dashed lines to
$N$(CO)/d$V=3\times10^{17} \cm2/\kms$. On each track, temperatures of
150, 100, 60, 30, 20 and 10 K are marked from left to right by small
open circles. }
\label{ratios}
\end{figure*}

Very approximate column densities may be calculated assuming optically
thin [CI] and $\13co$ emission in the high-temperature LTE limit, but
accurate results are obtained only by more detailed radiative transfer
calculations. Curves illustrating the possible physical condition of
gas clouds, shown in Fig.\,\ref{ratios}, were calculated with the
Leiden radiative transfer models (see
http://www.strw.leidenuniv.nl/\~\rm radtrans/).  The four panels
correspond to molecular gas densities ranging from $n(\h2) = 500 \cc$
to $n(\h2) = 10 000 \cc$. In each panel, mostly horizontal tracks mark
CO gradients $N$(CO)/d$V$ = 0.3, 1.0 and 3.0 $\times 10^{17}$
$\cm2/\kms$ for [C$^{\circ}$]/CO abundances of 0.1, 0.3, 1.0 and 3.0
respectively and kinetic temperatures ranging from 150 K (left) to 10
K (right) are marked. Lines of constant kinetic temperature would be
mostly vertical in these panels. The diagrams contain points
representing the observed galaxy centers as well as star-forming
regions (White $\&$ Sandell 1995; Israel $\&$ Baas, unpublished;
Bolatto et al. 2000) and the Milky Way center (Fixsen, Bennett $\&$
Mather 1999).

The star-forming PDRs have relatively low neutral carbon versus CO
abundances (C$^{\circ}$)/CO) $\approx$ 0.1--0.3). {\em Most galaxy
centers (especially the active ones) have significantly higher
abundances exceeding unity}, independent of the assumed gas 
parameters.  The diagonal distribution of galaxy centers roughly
follows lines of constant kinetic temperature.  The actual
temperatures $T_{\rm k}$ and molecular gas densities $n(\h2)$ cannot
be determined indepently. For instance, if we assume $n = 500 \cc$, we
find $T_{\rm kin} > 150$, whereas by assuming $n \geq 3000 \cc$ we
find more modest values $T_{\rm kin} = 30 - 60$ K similar to the dust
temperatures 33 K $\leq T_{\rm d} \leq$ 52 K found for these galaxy
centers by Smith $\&$ Harvey (1996). The only available direct
determination, for M 82 by Stutzki et al. (1997), yields $n \geq
10^{4} \cc$ and $T_{\rm k}$ = 50 -- 100 K, in very good agreement with
the above.  Where the molecular line emission in most galaxy centers
appears to arise from warm, dense gas (as opposed to either hot and
tenuous or cold and very dense gas), the centers of NGC~7331, M~51 and
NGC~4826 seem to be cooler independent of assumed gas density.

\section{[CI], [CII] and FIR intensities}

%
\begin{figure}
\unitlength1cm
\begin{center}
\begin{minipage}[b]{8.4cm}
\hspace{0.35cm}
\resizebox{6.74cm}{!}{\includegraphics*{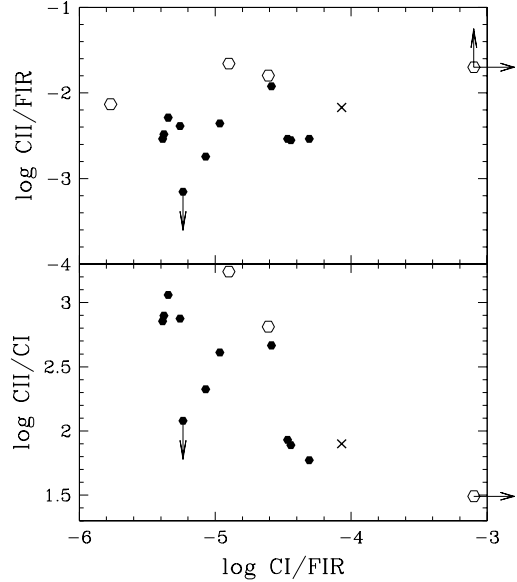}}
\end{minipage}
\caption[]{Ionized carbon [CII] line to far-infrared continuum (FIR) 
ratio (top) and [CII]/[CI] line ratio (bottom) as a function of 
neutral carbon [CI] line to FIR ratio. In both diagrams, the position 
of the Milky Way center is marked by a cross and the positions of 
Magellanic Cloud objects by open hexagons.}
\label{ciifir}
\end{center}
\end{figure}

Fig.\,\ref{ciifir} combines [CI] line intensities with those of
[CII] and the far-infrared continuum (for references, see Israel $\&$
Baas 2002). The results resemble those obtained by Gerin $\&$
Phillips (2000). There is no longer a clear distinction between the
various types of objects. The [CII]/FIR ratio increases with [CI]/FIR,
but the [CII]/[CI] ratio {\it decreases} as [CI]/FIR increases.  An
upper limit for [CII] places the merger galaxy NGC 660 in the same
diagram positions as the ultraluminous mergers Arp~220 and Mrk~231
observed by Gerin $\&$ Phillips. The highest [CII]/FIR ratios are
reached for PDR model gas densities $n = 10^{3} - 10^{4} \cc$ (Gerin
$\&$ Phillips 2000), but fully half of the ratios in
Fig.\,\ref{ciifir} are much higher than predicted by the PDR models.
In low-metallicity environments, this may be explained by longer mean
free pathlengths of energetic UV photons (Israel et al. 1996).
However, this is not a credible explanation for high-metallicity (see
Zaritsky, Kennicutt $\&$ Huchra 1994) galaxy centers.

Ideally, the observations should be modelled by physical parameters
varying as a function of location in a complex geometry. Practically,
we may attempt to approach such a reality by assuming the presence of
a limited number of distinct gas components. The analysis of
multitransition $\co$, $\13co$ and [CI] observations of half a dozen
galaxy centers (Israel $\&$ Baas 2003 and references therein) suggests
that, within the observational errors, good fits to the data can be
obtained by modelling with combinations of dense/cool and tenuous/warm
gas components. At the same time, however, the energy requirements of
keeping relatively large mass fractions at relatively high
temperatures, and the peculiar [CI]/[CII] ratios again suggest that
exploration of excitation models other than PDRs might be fruitful.

\end{article}
\end{document}